\documentclass[12pt]{article}
\usepackage{amsmath, amsthm, amsfonts}
\usepackage[margin=1 in]{geometry}
\usepackage{blindtext, xcolor}
\usepackage{algorithm}
\usepackage{algorithmicx}
\usepackage{algpseudocode}
\usepackage{apacite}
\usepackage{natbib}
\usepackage{appendix}
\usepackage{rotating}
\usepackage{hyperref}
\usepackage{float}
\usepackage{graphicx}
\usepackage{caption,subcaption}
\usepackage{cleveref}
\usepackage{arydshln}
\usepackage{authblk}
\usepackage{xr,bm}

\newtheorem{innercustomthm}{Theorem}
\newenvironment{customthm}[1]
  {\renewcommand\theinnercustomthm{#1}\innercustomthm}
  {\endinnercustomthm}
\newcommand{\be} {\begin{eqnarray*}}
\newcommand{\ee} {\end{eqnarray*}}

\theoremstyle{definition}

\def\*#1{\bm{#1}}

\title{Selecting a significance level in sequential testing procedures for community detection}
\author[1]{Riddhi Pratim Ghosh}
\author[2]{Ian Barnett}

\affil[1]{Department of Mathematics and Statistics, Bowling Green State University}
\affil[2]{Department of Biostatistics, University of Pennsylvania}
\date{}

\begin{document}
\maketitle

\begin{abstract}
While there have been numerous sequential algorithms developed to estimate community structure in networks, there is little available guidance and study of what significance level or stopping parameter to use in these sequential testing procedures. Most algorithms rely on prespecifiying the number of communities or use an arbitrary stopping rule. We provide a principled approach to selecting a nominal significance level for sequential community detection procedures by controlling the tolerance ratio, defined as the ratio of underfitting and overfitting probability of estimating the number of clusters in fitting a network. We introduce an algorithm for specifying this significance level from a user-specified tolerance ratio, and  demonstrate its utility with a sequential modularity maximization approach in a stochastic block model framework. We evaluate the performance of the proposed algorithm through extensive simulations and demonstrate its utility in controlling the tolerance ratio in single-cell RNA sequencing clustering by cell type and by clustering a congressional voting network.
\end{abstract}


%%%%%%%%%%%%%%%%%%%%%%%%%%%%%%%%%%%%%%%%%%%%%%%%%%%%%%
%%%%%%%%%%%%%%%%%%%%%%%%%%%%%%%%%%%%%%%%%%%%%%%%%%%%%%
\newpage
\section{Introduction}
In the last few decades, there has been an increasing interest among physicists, computer and social scientists to study network data. Identifying community structure in a networks has gained particular attention: the vertices in networks are often found to cluster into related groups where vertices within a community are more likely to be connected [see, e.g., \cite{newman2004finding}, \cite{newman2006modularity}]. The ability to detect such communities is crucial to understand
the relationship between the structure and function of networks, such as the modeling
of networks \citep{cheng2009triangular}, the evolution of networks \citep{zhang2008evolution,shen2010spectral}, the resilience of networks \citep{albert1999diameter, cheng2010bridgeness}, and
the capacity of networks \citep{zhang2007enhancing}. The stochastic block model \citep{holland1983stochastic} is a popular model for community structures in network data where edge probabilities between and within communities are constant conditional on community membership.

Many community detection methods begin with a null model of no community structure. Historically, the most common approach involving a null model is the use of a node partition score that is large when nodes within a partition are highly interconnected, relative to what is expected under the null of no structure \citep{newman2006modularity,fortunato2010community}. Many sequential community detection algorithms perform this task by first dividing the network into two communities, and subsequently subdividing each community hierarchically, ideally terminating when the true number of communities, $K$, has been reached.  One such algorithm that is widely used in literature is based on modularity maximization proposed by \citet{newman2006modularity} and its different variants including fast greedy modularity optimization \citep{clauset2004finding}, exhaustive modularity optimization via simulated annealing (\cite{guimera2004modularity}, \cite{massen2005identifying}, \cite{medus2005detection}, \cite{guimera2005functional}), fast modularity optimization \citep{blondel2008fast}.  Parallel community detection algorithms have garnered some attention over the last decade that modify existing algorithms to make them more suitable for the  analysis of large networks. \cite{riedy2011parallel} modified the agglomerative community detection algorithm by choosing multiple contraction edges simultaneously  as opposed to sequential contraction that is commonly done. \cite{yang2016comparative} compare several state-of-the-art algorithms on artificial networks in terms of accuracy and computing time.
\cite{que2015scalable} proposed a parallel community detection algorithm derived from Louvain modularity maximization method using a novel graph mapping and data representation. A hypothesis testing framework based on modularity-based community detection has been studied by \cite{zhang2017hypothesis} where they introduced a hypothesis testing procedure to determine the significance of the partitions obtained from maximizing the modularity function starting from a null model with no graph structure. However, this neglects the sequential nature of the test, and ignores correlations among test statistics which we incorporate in our approach. \cite{bickel2016hypothesis} provides an algorithm for finding the number of clusters
 in  a stochastic block  framework using the Tracy-Widom distribution as the limiting distribution of the highest eigenvalue of the adjacency matrix, and therefore is not suitable for the small or moderate sized networks. 
                  To make a sequential community detection algorithm effective, the significance level for rejecting the null hypothesis needs to be specified for each test given by $H_0: K=j$ community against $H_a: K>j$ starting with $j=1$ and incrementing $j$ over the integers until the test fails to reject $H_0$.
The standard practice of setting the significance level arbitrarily to to 0.05 or 0.01 has drawbacks because it is susceptible to multiple testing leading to increased Type I error due to the repeated sequential tests.

To circumvent the multiple testing problem in sequential community detection procedures, analogous to controlling family-wise error rate, specifying a nominal significance level accounting for multiple tests is necessary. We aim to instead control for the underfitting (overfitting) probability, defined as the probability that the estimated number of communities obtained by a sequential testing procedure is less than (greater than) the true number of communities $K$ present in the network.  Any given contexts specific tolerance for overfitting and underfitting probabilities ultimately dictates the nominal significance level that should be used. We address the problem of finding the nominal significance level and aim to provide an algorithm to determine it aligns with a context-specific user-specified \textit{tolerance ratio}, defined as the ratio of underfitting probability to overfitting probability in a generic sequential testing framework. Our algorithm hinges on finding a suitable estimate of the number of communities at a significance level that preserves the prespecified tolerance ratio.

The rest of this article is organized as follows. In Section 2, we first describe sequential community detection procedures and subsequently introduce our algorithm  to choose a significance level guided by a pre-specified tolerance ratio. In Section 3, we provide an example of our approach applied to Newman's modularity maximization for sequential community detection to select an appropriate significance level. Section 4 describes the performance of our algorithm through extensive simulation studies in stochastic block model frameworks. We derive appropriate significance levels in two real applications in Section 5. Finally Section 6 concludes with a discussion of limitations and next directions for our approach.

%%%%%%%%%%%%%%%%%%%%%%%%%%%%%%%%%%%%%%%%%%%%%%%%%%%
%%%%%%%%%%%%%%%%%%%%%%%%%%%%%%%%%%%%%%%%%%%%%%%%%%%%

%%%%%%%%%%%%%%%%%%%%%%%%%%%%%%%%%%%%%%%%%%%%%%%%%%%%
%%%%%%%%%%%%%%%%%%%%%%%%%%%%%%%%%%%%%%%%%%%%%%%%%%%%

\section{Sequential community detection}\label{community_detection_section}

In this section, we first describe a generalizable sequential testing procedure to detect the number of communities in a network. Secondly, we describe the estimation of the tolerance ratio by deriving the expressions of underfitting and overfitting probabilities using an estimate of the number of communities. This tolerance ratio estimate is a function of the nominal significance level, which we can then solve for to arrive at a desired prespecified level.

\subsection{Sequential testing procedure}
Assuming a network of size $n$, the sequential testing procedure can be described by the following hypotheses: 
\begin{align}\label{hypothesis_defn}
 & H_0:K=j, \;\;\;\;\;\ \text{ against } \;\;\;\;\ H_A:K>j,\\ \nonumber
\end{align}
for each integer $j \geq 1$ until a test fails to reject.

\subsection{Significance level from tolerance ratio}\label{algo_des}

A common problem faced in community detection is the choice of an appropriate significance level $\alpha$. Analogous to multiple testing problem \citep{benjamini1995controlling} where the goal is to control the  family-wise error rate (FWER) through some procedures such as Bonferroni correction, Tukey's range test etc.,  we focus on sequential community detection algorithms, where tests of the null hypothesis $H_0: K=j$ against the alternative $H_A: K>j$ are performed sequentially for $j=1,2,\ldots$ until a test fails to reject $H_0$. We let $p(j)$ be the p-value of the $j$th such test $T(j;\alpha)$ defined as:
  \[
    T(j;\alpha) = \left\{\begin{array}{lr}
        1 & \text{  for } p(j)\leq\alpha\\
        0 & \text{  for } p(j)>\alpha
        \end{array}\right.
  \]
Using this sequential procedure, the estimated number of communities is:
\begin{align}\label{K_hat_eqn}
& \hat{K}(\alpha) = \inf\{k \in \mathbb{N} : T(k+1;\alpha)=0\},\\ \nonumber
\end{align}

 is a non-decreasing (step) function of $\alpha$.

 We define the underfitting probability to be $\mbox{pr}(\hat{K}(\alpha) <K)=\eta_{u}$ and the overfitting probability to be $\mbox{pr}(\hat{K}(\alpha) >K)=\eta_{o}$. 
The \textit{tolerance ratio} is defined as $\gamma=\eta_u/\eta_o$, where $K$ is the true number of communities. One can note that $\gamma \in [0,\infty)$. In particular, $\gamma=1$ implies underfitting and overfitting probabilities are equally likely. For unknown $K$, this also suggests one approach to estimate $K$ that is independent of $\alpha$: select $\hat{K}$ to be the value of $K$ that results from the widest subinterval of $\alpha$ in $[0,1]$. We call this \textit{$\alpha$-free} estimator $K^{*}$.

We propose the following iterative procedure to identify the correct marginal significance level $\alpha$ to use from the user-specified tolerance ratio $\gamma$.

\hspace{0.2in}

\hrule 

\hspace{0.00001in}

{\bf{Input:}} The original or estimated adjacency matrix $A$ of a graph and  user-specified tolerance $\gamma$\\
\hrule

\hspace{0.0000001in}
\begin{enumerate}

\item \underline{For a given $\alpha$, perform sequential community detection to obtain $\hat{K}(\alpha)$}: For each $k \in \{1,2,\cdots,n\}$, we simulate SBM of size $n$ and cluster $k$ and use $\mathcal{B}$ bootstrap samples to compute the test statistic at  $k$-th stage . This will give us empirical null distribution of the test statistic. Next we compare it with the observed value of the test statistic and find $\hat{K}(\alpha)$ according to \eqref{K_hat_eqn}.

\item \underline{Determining $K^{*}$}: $\hat{K}(\alpha)$ is a non-decreasing step function of $\alpha$ which can take integer values between 1 and $n$. Let $\alpha_1<\alpha_2<...<\alpha_m$ denote the values of $\alpha$ in $[0,1]$ which yield distinct values $\hat{K}(\alpha_1)<\hat{K}(\alpha_2)<...<\hat{K}(\alpha_m)$, where $1\leq m\leq n$. Let $I_j=\{\alpha \in [0,1]: \hat{K}(\alpha)=\hat{K}(\alpha_j)\}$ for $j=1,2,...,m$, and $M=\underset{1\leq j \leq m}{\text{argmax}}\text{ length}(I_j)$. Define $K^*=\hat{K}(\alpha_{M})$. In other words, $K^*$ is the longest step when $\hat{K}(\alpha)$ is plotted with respect to $\alpha$.

\item \underline{For a given $\alpha$, compute the tolerance ratio $\gamma(\alpha)$}:
$$\gamma(\alpha)=\frac{\frac{1}{\mathcal{B}}\sum_{b=1}^\mathcal{B}I_{\{\hat{K}^{(b)}(\alpha)<K^{*}\}}}{\frac{1}{\mathcal{B}}\sum_{b=1}^\mathcal{B}I_{\{\hat{K}^{(b)}(\alpha)>K^{*})\}}}$$ using $\mathcal{B}$ bootstrap samples.

\item Finally, find $\alpha \in [0,1]$ such that $|\gamma(\alpha)-\gamma|$ is the minimum.

\end{enumerate}

\hrule

\vspace{.3in}

The above algorithm yields an $\alpha$ which can be repeatedly seeded back into Step 1 until convergence in $\alpha$ is achieved.

Below we present a brief proof of the convergence of the algorithm which assumes the Lipschitz condition on $\gamma$, and exploits some key characteristics about the change of underfitting probability with respect to $\alpha$.

\begin{customthm}{1}\label{eight}
Suppose the target value of the significance is $\alpha_0$ that corresponds to the tolerance ratio $\gamma_0$. Also assume that the function $\gamma(\alpha)$ satisfies the Lipschitz condition 
$$|\gamma(\alpha)-\gamma(\alpha^*)|\leq c|\alpha-\alpha^*|,$$
where $c>0$ is the Lipschitz constant,  $\alpha,\alpha^* \in (0,1)$, and $\mathcal{B}$ tends to $\infty$. Then if we terminate the algorithm with precision $\epsilon$ for the significance $\alpha$, then the precision of the tolerance ratio  is $c\epsilon$.
\end{customthm}

\begin{proof}
First note that $\gamma(\alpha)=P(\hat{K}(\alpha)<K^*)/P(\hat{K}(\alpha)>K^*)$ is an increasing function of the underfitting probability $P(\hat{K}(\alpha)<K^*)$, which in turn is a decreasing function of $\alpha$. This implies $\gamma(\alpha)$ is a decreasing function of $\alpha$. So, there exists a $c$ (could be very large) so that the Lipschitz condition holds. Therefore, if we use the precision $\epsilon$ for $\alpha$, the precision for $\gamma$ is $c\epsilon$.

\end{proof}

Two remarks are in order.\\

\noindent
\textit{Remark 1.} In \Cref{eight}, we assume that $\gamma(\alpha)$ is a continuous decreasing function of $\alpha$. However, in Step 3 of our algorithm, $\gamma(\alpha)$ is guaranteed to be a non-increasing step function of $\alpha$ because we are estimating it empirically and $\hat{K}^{(b)}(\alpha)$ can take finitely many values in $\{1, 2, ...,n\}$. Therefore the difference $|\gamma(\alpha)-\gamma(\alpha^*)|$ can range over the entire real line, taking only finite values. The difference $|\alpha-\alpha^*|$ can range in the interval $(0,1)$. It is instructive to note that when the difference in $\alpha$ is zero or small, the corresponding difference in $\gamma$ is also zero. Therefore, one can always pick $c>0$ so that the Lipschitz condition is satisfied.\\

\noindent
\textit{Remark 2.} Our algorithm takes the user-specified tolerance as input and is expected to return a significance level is close to the tolerance ratio as possible. Like other iterative algorithms, we need some pre-defined precision that dictates the stopping criterion. This task is accomplished by fixing $\epsilon$ beforehand so that our algorithm returns a significance that lies within the $\epsilon$- neighborhood of the optimal significance. \Cref{eight} provides the stopping criterion in Step 4 based on the precision of $\alpha$. \\

 %%%%%%%%%%%%%%%%%%%%%%%%%%%%%%%%%%%%%%%%%%%%%%%%%%%%

\section{Example sequential community detection algorithm}

While our approach for identifying an $\alpha$ that corresponds with a prespecfied tolerance ratio is agnostic to which sequential community detection algorithm is used, we detail one example use case here. Aside various community detection algorithms such as  spectral clustering \citep{white2005spectral, zhang2007identification}, random walks \citep{pons2005computing}, a popular approach to community detection is based on the idea of optimizing \textit{modularity}. Modularity metrics were introduced by \cite{newman2004finding}, and the idea of detecting communities by optimizing a modularity function
was proposed by \cite{newman2004fast} Nowadays, there are many
variants of the modularity-based community detection approach to deal with directed or
weighted networks \citep{leicht2008community}. 
Also, some variants of the modularity-based community detection approach use modularity functions with a somewhat modified mathematical structure \citep{reichardt2006statistical, waltman2010unified, traag2011narrow}.

Here we revisit Newman's sequential algorithm \citep{newman2006modularity} of community detection which begins by first dividing the network into two communities  and then subdividing into further communities by maximizing additional modularity; and we implement our approach to selecting an appropriate significance level in this context.

For a network with $n$ vertices, let $A$ denote the $n \times n$ adjacency matrix and $s=(s_1,s_2,\cdots,s_n)^{\top}$ $\in \{-1,1\}^n$ where $s_i=1$ if the $i$-th vertex belongs to group 1 and -1 otherwise. Let $k_i$ denote the degree of vertex $i$ and $m=\sum_{i=1}^{n}k_i/2$ be the total number of edges in the network. Then the modularity of the network is defined as 
\begin{align}\label{modularity_defn}
 & Q=\frac{1}{4m}s^{\top}Bs,
\end{align}
where the matrix $B=(B_{uv})$ is defined as $B_{uv}=A_{uv}-\frac{k_uk_v}{2m}$, a symmetric matrix of order $n$.

Let $u_1, u_2, \cdots, u_n$ be the eigenvectors of $B$ corresponding to the eigenvalues $\lambda_1 \geq \lambda_2 \geq \ldots \geq \lambda_n$. Then $Q$ in \eqref{modularity_defn} is maximized if $s_i=1$ if the corresponding element in $u_1$ is positive and $s_i=-1$ otherwise rendering a network divided into two communities. 

For further dividing a group $j$ of size $n_j$, the additional contribution to the modularity is 
\begin{align} \label{addln_modularity_defn}
 & \delta Q_j=\frac{1}{4m}s^{\top}B^{(j)}s    
\end{align}
is maximized in the similar way for $Q$ in \eqref{modularity_defn}, where
$B_{uv}^{(j)}=B_{uv}-\delta_{uv}\sum_{l \in j}B_{ul}$, and $\delta_{uv}$ is the Kronecker $\delta$-symbol.

If the total modularity of the network after splitting the network into $j$ communities is $Q^{(j)}$, then the gain in the modularity is defined by $\Delta Q^{(j)}=Q^{(j+1)} - Q^{(j)}$. Again, while we use this quantity $\Delta Q^{(j)}$ as our test statistic for the $j$th step ($H_0: K=j$ vs $H_A: k>j$), we stress that any sequential community detection algorithm can be adopted to this framework.

%%%%%%%%%%%%%%%%%%%%%%%%%%%%%%%%%%%%%%%%%%%%%%%%%%%%%%%

\section{Simulations}
We perform extensive simulation study in various directions to assess the performance of the proposed algorithm. In each set-up, networks of size $n$ and $2n$ are simulated through SBM with $K_0$ number of balanced communities of size $n/K_0$. We vary $n \in \{100, 200\}$ corresponding to $K_0 = 5 ,10$ respectively for symmetric edge probability matrix $P$ of dimension $K_0$ of the form
$$P=2\epsilon \bm{I}_k+ (0.5-\epsilon)\bm{1}_k\bm{1}_k^{\top},$$
so that the diagonal and off-diagonal entries of $P$ are $0.5+\epsilon$ and $0.5-\epsilon$ respectively implying that the difference between edge probability within and between community is $2\epsilon$. We vary $\epsilon=0.195, 0.010$ to represent two cases of (S) strong and (W) weak community structure, respectively.

\subsection{Estimated number of communities ($\hat{K}_{\alpha}$)}
For a fixed $\alpha$, we simulate 1000 parametric bootstrap sample values of the null test statistic and calculate p-values by comparing them with the observed test statistic value. We start from the number of communities  $K=1$, and proceed by incrementing $K$until the p-value is greater than $\alpha$. We replicate the procedure 100 times and finally report the value of the estimated number of communities $\hat{K}_{\alpha}$ by taking the mode of the 100 replications.

Next, we vary $\alpha \in \{0.01,0.05,0.10,0.20\}$ and the corresponding $\hat{K}_{\alpha}$s are reported in  \Cref{Khatalpha}. Further, the estimates of  $\text{pr}(\hat{K}_{\alpha}=K_0)$ are reported in \Cref{PofKhatalpha} by taking the proportion of times $\hat{K}_{\alpha}$ is equal to $K_0$ over 100 replications. 
 It is instructive to note that entries in \Cref{Khatalpha} is less than the significance level $\alpha$. It can be shown by straightforward calculation that for a given $\alpha$, $\text{pr}(\hat{K}_{\alpha}=K_0) < \alpha$.

One can note that in the presence of  strong differences in communities, estimated communities are close to the true number for $\alpha=0.01,...,0.2$. For weak signals, the number of communities is under estimated for the aforementioned $\alpha$. However, the number of communities is over estimated for larger value of the significance level. This indicates that the choice of $\alpha$ can greatly influence $\hat{K}$, which provides further incentivize for developing a rigorous approach to selecting an appropriate $\alpha$.

\begin{table}[ht]
 \caption{Mode of 100 independent replications ($\hat{K}_{\alpha}$), and proportion of times true number of communities  correctly estimated \big($\text{pr}(\hat{K}=K_0)$\big) for different choice of $\alpha$, $P$, and $n$.}
 \label{Khatalpha}
 \begin{center}
\begin{tabular}{ c|c|c|c|c|c}
 \hline
 $K_0$  & Signal & $\alpha=0.01$ &  $\alpha=0.05$ & $\alpha=0.1$ & $\alpha=0.2$\\
 \hline
 5  &  S & 5 (0.82) &  5 (0.85) & 5 (0.85) & 6 (0.45) \\
  & W & 3 (0.10) & 3 (0.15) & 4 (0.35) & 4 (0.20)\\
 %&  W & 1 (0.00) &1 (0.00) & 2 (0.00) &2 (0.00)\\
\hline
  10 &  S & 8 (0.25)&  8 (0.35)& 10 (0.40)& 10 (0.55)\\
  & W & 2 (0.00)& 2 (0.00)& 3 (0.00) & 3 (0.00) \\
  % &  W & 1 (0)& 1 (0)& 1 (0) & 1 (0)\\
\hline
\end{tabular}
\end{center}
\end{table}

\begin{table}[!ht]
 \caption{Mode of 50 independent replications ($\hat{K}_{\alpha}$), and proportion of times true number of communities  correctly estimated \big($\text{pr}(\hat{K}=K_0)$\big) for different choice of $\alpha$, $P$, and $2n$.}
 \label{PofKhatalpha}
 \begin{center}
\begin{tabular}{ c|c|c|c|c|c}
 \hline
 $K_0$  &  Signal & $\alpha=0.01$ &  $\alpha=0.05$ & $\alpha=0.1$ & $\alpha=0.2$\\
 \hline
 5 & S  &  5 (0.85) & 5 (0.88) & 5 (0.89) & 5 (0.55) \\
  & W &  4 (0.30) & 5 (0.50) & 5 (0.58) & 5 (0.45) \\
%  & W &   &   &  &  \\
\hline
 10 & S & 9 (0.49) &  10 (0.75) & 10 (0.85) & 10 (0.70)\\
   & W & 3 (0.00) & 6 (0.00) & 6 (0.05) & 7 (0.10)  \\
  % & W &   &   &   &  \\
\hline
\end{tabular}
\end{center}
\end{table}

\subsection{Choice of significance level ($\alpha$)}

In each simulation set-up, we use $1000$ bootstrap samples for a wide range of $\alpha$ (typically in the range $[0.001,0.5]$) and store the values of $\hat{\gamma}(\alpha)$ according to Step 3 of the algorithm. 

We consider the value of tolerance ratio  $\eta_u=0.5 ,1, 2$ corresponding to the cases where underfitting probability is half, equal, and twice of overfitting probability. In each case, we find the $\hat{\gamma}(\alpha)$ such that $|\hat{\gamma}(\alpha)-\gamma|$ is the minimum among the stored values, and report the corresponding value of $\alpha$ in \Cref{underfitting.table1}. One can note that as $\gamma$ is increasing (i.e., the overfitting probability is increasing relative to the underfitting probability), the significance level decreases. This is consistent with the fact that for a smaller value of $\alpha$, the test is getting accepted at an early step than a larger value of $\alpha$. 
%\iffalse

\begin{table}[!ht]
 \caption{Choice of $\alpha$ for different choices of  tolerance ratio $\gamma$ and network size $n(n'=2n)$ }
 \label{underfitting.table1} 
 \begin{center}
\begin{tabular}{ c|c|c|c|c}
 \hline
 $K_0$  &  Signal &  $\gamma=1/2$ & $\gamma=1$ & $\gamma=2$ \\
 \hline
 5 & S &  0.06 (0.06)  &  0.01 (0.02) & 0.005 (0.006)   \\
  & W &  0.10 (0.09)  & 0.05 (0.04) & 0.02 (0.01) \\
\hline
  10 & S    &  0.07 (0.06) &    0.01 (0.01) &  0.005 (0.006)   \\
  & W   &  0.10 (0.10) &  0.05 (0.04) & 0.01 (0.01)  \\
 \hline
\end{tabular}
\end{center}
\end{table}

\vspace{3in}

\newpage

%%%%%%%%%%%%%%%%%%%%%%%%%%%%%%%%%%%%%%%%%%%%%%%%%%%%%%%

\section{Real data analysis}

\subsection{Single cell RNA (scRNA-seq) data}

\begin{figure}[h!]
\centering
\includegraphics[width=160mm]{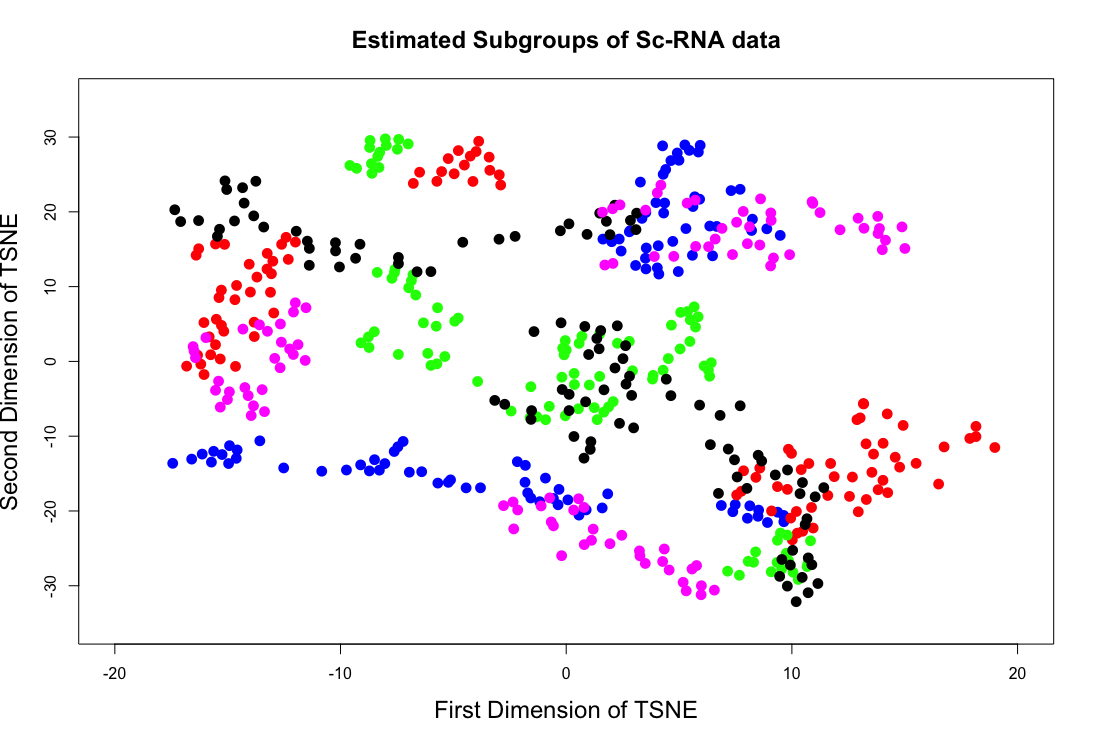}
\caption{Virtual representation of the estimated number of clusters of the analyzed scRNA-seq data of human retina cells in the t-SNE plot obtained by selecting Seurat classified cell types namely: Astrocytes, Endothelium, Ganglion, Horizaontal, Pericytes in an equal manner of roughly 100 cells per cell type.} 
\label{method}
\end{figure}

We apply our algorithm to the scRNA-seq data generated from the retina cells of two healthy
adult donors using the 10X $\text{Genomics Chromium}^{\text{TM}}$ system. We should expect some clustering by cell type in networks derived from this data. Detailed preprocessing
and donor characteristics of the scRNA-seq data can be found
in \cite{lyu2019integrative}.  The data consists of 33694 genes sequenced over 92385 cells. The sequencing data were initially analyzed with
R package \textit{Seurat} \citep{satija2015spatial} and each of the cells was identified
as a particular cell-type. The virtual representation of the data in the t-SNE plot is given in \Cref{method}.

Among different clusters in \textit{Seurat}, we consider the data pertaining to five hierarchical clusters: ``Astrocytes", ``Endothelium", ``Ganglion",  ``Horizontal", ``Pericytes". Before we perform the analysis, we process the data in three steps. First, genes whose variability was less than the 50th quantile are filtered out, and then cells whose total cell counts across all genes are less than 500 and greater than 2500 are also filtered out. Second, we compute the normalized score (row wise) and perform a log
transformation ($log_2(1+x/10000))$ as done in \cite{booeshaghi2021normalization} to
convert the data into a continuous scale. The rationale behind such a transformation is that  that different genes have different variances implying that genes that are highly expressed will have high variance whereas the genes that are barely expressed at all, will have almost zero variance.  The transformed data is now 
used to compute correlations between the cells. Finally, for each cluster, we randomly select 100 cells ensuring that the within and between cluster correlations do not differ by more than 0.1 from those of the composite data. We use the correlation threshold ($\tau$) to
construct an adjacency matrix $A$, and vary $\tau \in \{0.3, 0.5, 0.7\}$, and report the significance level along with estimated number of communities in \Cref{underfitting.table.sc_RNA}. We observe that estimated number of communities is larger as we increase the value of $\tau$ which gives rise to a denser network. The significance level ($\alpha$) ranges over $[0.01,0.05]$ depending on the tolerance ratio. Also, for each choice of $\tau$, the estimated number of communities is increasing with $\alpha$.

It can also be noted from \Cref{underfitting.table.sc_RNA} that different values of $\alpha$ lead to different number of estimated communities. If one were to arbitrarily pick $\alpha$ as, say, 0.05, this choice can have a large impact on the analysis. For example, corresponding to $\tau=0.3$, $\alpha$ changes from 0.05 to 0.01  leading to different value of $\hat{K}$. Thus the choice of $\alpha$ is an impactful decision, and the tolerance ratio presents an intuitive measure that allows the practitioner to place a value of overfitting relative to underfitting when performing community detection.

\begin{table}[!ht]
 \caption{Choice of $\alpha$ and corresponding estimated number of communities $\hat{K}_{\alpha}$ for different values of tolerance ratio  $\eta$  across various choices of correlation threshold $\tau$ for the scRNA-seq data.}
 \label{underfitting.table.sc_RNA} 
 \begin{center}
\begin{tabular}{ c|ccc|ccc|ccc}
\hline
 Correlation threshold ($\tau$) &  & 0.3 & & & 0.5 & & & 0.7 & \\
\hline
  Tolerance ratio ($\eta$) & 0.5 &  1 &  2  & 0.5 &  1 & 2 & 0.5 &  1 & 2  \\
 %\hline
 Significance level ($\alpha$) & 0.05 & 0.03  &   0.01  &  0.04  &  0.04 &   0.01 &  0.05  & 0.03  &  0.02  \\
% \hline
  Estimated $\#$ communities ($\hat{K}_{\alpha}$) & 7 &  7 &  6   &   9  &    8 &   7  &   9   &  9   &  7  \\
 \hline
\end{tabular}
\end{center}
\end{table}

\subsection{United States House Votes 1984 (USHV) data}

In this example, we consider a data set of 267 democrats and 167 republican congressmen who has voted in 16 issues in 1984 in the United States of America. The data contains yes/no answer for each congressman on 16 different questions with some missing values. After removing the congressman who has not voted in more than three of the sixteen questions, the data is represented by a $417 \times 17$ matrix where the first column represents the political affiliation-republican and democrat. The adjacency matrix $A$ is calculated by thresholding the correlations among congressmen by $\tau$, i.e., if the correlation of voting between two congressmen is as high as $\tau$, we assume they are connected by an edge and hence the corresponding entry of the adjacency matrix 1, and 0 otherwise. Finally we vary $\tau \in \{0.3, 0.5, 0.7\}$. It is instructive to note that smaller values of $\tau$ leads to a more dense network.

In this data, the number of distinct communities is not expected to go below 2 because of the two party affiliations. However, in our analysis, the estimated number of communities varies in $\{3, 4, 5\}$ (depending on the desired tolerance ratio) implying potential further subdivisions among political parties. Here too, the significance level has a large impact on the analysis. For example, when $\tau=0.7$, changing $\alpha$ from 0.06 to 0.04 drops the number of communities from 5 to 4. Therefore, a judicious choice of the significance level is necessary, and the tolerance ratio again provides a means of guiding this choice in an intuitive manner.

\begin{table}[!ht]
 \caption{Choice of $\alpha$ and corresponding estimated number of communities $\hat{K}_{\alpha}$ for different values of tolerance ratio  $\eta$  across various choices of correlation threshold $\tau$ for the USVA data}
 \label{underfitting.table} 
 \begin{center}
\begin{tabular}{ c|ccc|ccc|ccc}
\hline
 Correlation threshold ($\tau$) &  & 0.3 & & & 0.5 & & & 0.7 & \\
\hline
  Tolerance ratio ($\eta$) & 0.5 &  1 &  2  & 0.5 &  1 & 2 & 0.5 &  1 & 2  \\
 %\hline
 Significance level ($\alpha$) & 0.07 & 0.05  &   0.02  &  0.06  &  0.05 &   0.01 &  0.06  & 0.04  &  0.01  \\
% \hline
  Estimated $\#$ communities ($\hat{K}_{\alpha}$) & 3 &  3 &  3   &   4  &    3 &   3  &   5   &  4   &  4  \\
 \hline
\end{tabular}
\end{center}
\end{table}

\newpage

\section{Discussion}

We have proposed an algorithm to provide guidance to the practitioner in order to obtain a nominal significance level that matches their desired balance between overfitting and underfitting probabilities. Traditional approaches to estimate the number of communities often arbitrarily set the significance level, and the tolerance ratio presents a intuitive alternative. To construct the test statistic in a sequential testing framework, we demonstrated the approach with Newman's modularity maximization method, although the procedure is general and can be applied equally to any sequential community detection approach. 

Although here we have assumed a stochastic block model, a feasible extension of this approach would be to apply it to dynamic stochastic block models \cite{matias2015statistical} in order to allow a time varying network structure. It is instructive to note that we proposed the solution using the sequential tests, and implemented the algorithm via bootstrap due to the lack of the analytic expression of the test statistic. A potential bottleneck that the proposed algorithm will face is when the network size is very large because bootstrapping will be computationally expensive. However, in case an analytic expression of the test statistic is available in closed form, the algorithm can be adapted trivially to use it in place of bootstrapping. This would further increase algorithmic stability by removing stochasticity introduced through the bootstrap.
 
\section*{Acknowledgements}
The authors (RG \& IB) would like to thank Mingyao Li, Professor of Biostatistics at the University of Pennsylvania for the scRNA-seq data. IB is supported by R01MH116884.

\medskip
%Bibliographystyle{IEEEtran}
%Bibliography{reference}{}
%Bibliographystyle{plainnat}
\bibliographystyle{apacite}

\bibliography{reference}

\begin{thebibliography}{}

\bibitem [\protect \citeauthoryear {%
Albert%
, Jeong%
\BCBL {}\ \BBA {} Barab{\'a}si%
}{%
Albert%
\ \protect \BOthers {.}}{%
{\protect \APACyear {1999}}%
}]{%
albert1999diameter}
\APACinsertmetastar {%
albert1999diameter}%
\begin{APACrefauthors}%
Albert, R.%
, Jeong, H.%
\BCBL {}\ \BBA {} Barab{\'a}si, A\BHBI L.%
\end{APACrefauthors}%
\unskip\
\newblock
\APACrefYearMonthDay{1999}{}{}.
\newblock
{\BBOQ}\APACrefatitle {Diameter of the world-wide web} {Diameter of the
  world-wide web}.{\BBCQ}
\newblock
\APACjournalVolNumPages{Nature}{401}{6749}{130--131}.
\PrintBackRefs{\CurrentBib}

\bibitem [\protect \citeauthoryear {%
Benjamini%
\ \BBA {} Hochberg%
}{%
Benjamini%
\ \BBA {} Hochberg%
}{%
{\protect \APACyear {1995}}%
}]{%
benjamini1995controlling}
\APACinsertmetastar {%
benjamini1995controlling}%
\begin{APACrefauthors}%
Benjamini, Y.%
\BCBT {}\ \BBA {} Hochberg, Y.%
\end{APACrefauthors}%
\unskip\
\newblock
\APACrefYearMonthDay{1995}{}{}.
\newblock
{\BBOQ}\APACrefatitle {Controlling the false discovery rate: a practical and
  powerful approach to multiple testing} {Controlling the false discovery rate:
  a practical and powerful approach to multiple testing}.{\BBCQ}
\newblock
\APACjournalVolNumPages{Journal of the Royal Statistical Society: Series B
  (Methodological)}{57}{1}{289--300}.
\PrintBackRefs{\CurrentBib}

\bibitem [\protect \citeauthoryear {%
Bickel%
\ \BBA {} Sarkar%
}{%
Bickel%
\ \BBA {} Sarkar%
}{%
{\protect \APACyear {2016}}%
}]{%
bickel2016hypothesis}
\APACinsertmetastar {%
bickel2016hypothesis}%
\begin{APACrefauthors}%
Bickel, P\BPBI J.%
\BCBT {}\ \BBA {} Sarkar, P.%
\end{APACrefauthors}%
\unskip\
\newblock
\APACrefYearMonthDay{2016}{}{}.
\newblock
{\BBOQ}\APACrefatitle {Hypothesis testing for automated community detection in
  networks} {Hypothesis testing for automated community detection in
  networks}.{\BBCQ}
\newblock
\APACjournalVolNumPages{Journal of the Royal Statistical Society: Series B
  (Statistical Methodology)}{78}{1}{253--273}.
\PrintBackRefs{\CurrentBib}

\bibitem [\protect \citeauthoryear {%
Blondel%
, Guillaume%
, Lambiotte%
\BCBL {}\ \BBA {} Lefebvre%
}{%
Blondel%
\ \protect \BOthers {.}}{%
{\protect \APACyear {2008}}%
}]{%
blondel2008fast}
\APACinsertmetastar {%
blondel2008fast}%
\begin{APACrefauthors}%
Blondel, V\BPBI D.%
, Guillaume, J\BHBI L.%
, Lambiotte, R.%
\BCBL {}\ \BBA {} Lefebvre, E.%
\end{APACrefauthors}%
\unskip\
\newblock
\APACrefYearMonthDay{2008}{}{}.
\newblock
{\BBOQ}\APACrefatitle {Fast unfolding of communities in large networks} {Fast
  unfolding of communities in large networks}.{\BBCQ}
\newblock
\APACjournalVolNumPages{Journal of Statistical Mechanics: Theory and
  Experiment}{2008}{10}{P10008}.
\PrintBackRefs{\CurrentBib}

\bibitem [\protect \citeauthoryear {%
Booeshaghi%
\ \BBA {} Pachter%
}{%
Booeshaghi%
\ \BBA {} Pachter%
}{%
{\protect \APACyear {2021}}%
}]{%
booeshaghi2021normalization}
\APACinsertmetastar {%
booeshaghi2021normalization}%
\begin{APACrefauthors}%
Booeshaghi, A\BPBI S.%
\BCBT {}\ \BBA {} Pachter, L.%
\end{APACrefauthors}%
\unskip\
\newblock
\APACrefYearMonthDay{2021}{}{}.
\newblock
{\BBOQ}\APACrefatitle {Normalization of single-cell RNA-seq counts by log (x+
  1) or log (1+ x)} {Normalization of single-cell rna-seq counts by log (x+ 1)
  or log (1+ x)}.{\BBCQ}
\newblock
\APACjournalVolNumPages{Bioinformatics}{37}{15}{2223--2224}.
\PrintBackRefs{\CurrentBib}

\bibitem [\protect \citeauthoryear {%
Cheng%
, Ren%
, Shen%
, Zhang%
\BCBL {}\ \BBA {} Zhou%
}{%
Cheng%
\ \protect \BOthers {.}}{%
{\protect \APACyear {2010}}%
}]{%
cheng2010bridgeness}
\APACinsertmetastar {%
cheng2010bridgeness}%
\begin{APACrefauthors}%
Cheng, X\BHBI Q.%
, Ren, F\BHBI X.%
, Shen, H\BHBI W.%
, Zhang, Z\BHBI K.%
\BCBL {}\ \BBA {} Zhou, T.%
\end{APACrefauthors}%
\unskip\
\newblock
\APACrefYearMonthDay{2010}{}{}.
\newblock
{\BBOQ}\APACrefatitle {Bridgeness: a local index on edge significance in
  maintaining global connectivity} {Bridgeness: a local index on edge
  significance in maintaining global connectivity}.{\BBCQ}
\newblock
\APACjournalVolNumPages{Journal of Statistical Mechanics: Theory and
  Experiment}{2010}{10}{P10011}.
\PrintBackRefs{\CurrentBib}

\bibitem [\protect \citeauthoryear {%
Cheng%
, Ren%
, Zhou%
\BCBL {}\ \BBA {} Hu%
}{%
Cheng%
\ \protect \BOthers {.}}{%
{\protect \APACyear {2009}}%
}]{%
cheng2009triangular}
\APACinsertmetastar {%
cheng2009triangular}%
\begin{APACrefauthors}%
Cheng, X\BHBI Q.%
, Ren, F\BHBI X.%
, Zhou, S.%
\BCBL {}\ \BBA {} Hu, M\BHBI B.%
\end{APACrefauthors}%
\unskip\
\newblock
\APACrefYearMonthDay{2009}{}{}.
\newblock
{\BBOQ}\APACrefatitle {Triangular clustering in document networks} {Triangular
  clustering in document networks}.{\BBCQ}
\newblock
\APACjournalVolNumPages{New Journal of Physics}{11}{3}{033019}.
\PrintBackRefs{\CurrentBib}

\bibitem [\protect \citeauthoryear {%
Clauset%
, Newman%
\BCBL {}\ \BBA {} Moore%
}{%
Clauset%
\ \protect \BOthers {.}}{%
{\protect \APACyear {2004}}%
}]{%
clauset2004finding}
\APACinsertmetastar {%
clauset2004finding}%
\begin{APACrefauthors}%
Clauset, A.%
, Newman, M\BPBI E.%
\BCBL {}\ \BBA {} Moore, C.%
\end{APACrefauthors}%
\unskip\
\newblock
\APACrefYearMonthDay{2004}{}{}.
\newblock
{\BBOQ}\APACrefatitle {Finding community structure in very large networks}
  {Finding community structure in very large networks}.{\BBCQ}
\newblock
\APACjournalVolNumPages{Physical Review E}{70}{6}{066111}.
\PrintBackRefs{\CurrentBib}

\bibitem [\protect \citeauthoryear {%
Fortunato%
}{%
Fortunato%
}{%
{\protect \APACyear {2010}}%
}]{%
fortunato2010community}
\APACinsertmetastar {%
fortunato2010community}%
\begin{APACrefauthors}%
Fortunato, S.%
\end{APACrefauthors}%
\unskip\
\newblock
\APACrefYearMonthDay{2010}{}{}.
\newblock
{\BBOQ}\APACrefatitle {Community detection in graphs} {Community detection in
  graphs}.{\BBCQ}
\newblock
\APACjournalVolNumPages{Physics Reports}{486}{3-5}{75--174}.
\PrintBackRefs{\CurrentBib}

\bibitem [\protect \citeauthoryear {%
Guimera%
\ \BBA {} Amaral%
}{%
Guimera%
\ \BBA {} Amaral%
}{%
{\protect \APACyear {2005}}%
}]{%
guimera2005functional}
\APACinsertmetastar {%
guimera2005functional}%
\begin{APACrefauthors}%
Guimera, R.%
\BCBT {}\ \BBA {} Amaral, L\BPBI A\BPBI N.%
\end{APACrefauthors}%
\unskip\
\newblock
\APACrefYearMonthDay{2005}{}{}.
\newblock
{\BBOQ}\APACrefatitle {Functional cartography of complex metabolic networks}
  {Functional cartography of complex metabolic networks}.{\BBCQ}
\newblock
\APACjournalVolNumPages{Nature}{433}{7028}{895--900}.
\PrintBackRefs{\CurrentBib}

\bibitem [\protect \citeauthoryear {%
Guimera%
, Sales-Pardo%
\BCBL {}\ \BBA {} Amaral%
}{%
Guimera%
\ \protect \BOthers {.}}{%
{\protect \APACyear {2004}}%
}]{%
guimera2004modularity}
\APACinsertmetastar {%
guimera2004modularity}%
\begin{APACrefauthors}%
Guimera, R.%
, Sales-Pardo, M.%
\BCBL {}\ \BBA {} Amaral, L\BPBI A\BPBI N.%
\end{APACrefauthors}%
\unskip\
\newblock
\APACrefYearMonthDay{2004}{}{}.
\newblock
{\BBOQ}\APACrefatitle {Modularity from fluctuations in random graphs and
  complex networks} {Modularity from fluctuations in random graphs and complex
  networks}.{\BBCQ}
\newblock
\APACjournalVolNumPages{Physical Review E}{70}{2}{025101}.
\PrintBackRefs{\CurrentBib}

\bibitem [\protect \citeauthoryear {%
Holland%
, Laskey%
\BCBL {}\ \BBA {} Leinhardt%
}{%
Holland%
\ \protect \BOthers {.}}{%
{\protect \APACyear {1983}}%
}]{%
holland1983stochastic}
\APACinsertmetastar {%
holland1983stochastic}%
\begin{APACrefauthors}%
Holland, P\BPBI W.%
, Laskey, K\BPBI B.%
\BCBL {}\ \BBA {} Leinhardt, S.%
\end{APACrefauthors}%
\unskip\
\newblock
\APACrefYearMonthDay{1983}{}{}.
\newblock
{\BBOQ}\APACrefatitle {Stochastic blockmodels: First steps} {Stochastic
  blockmodels: First steps}.{\BBCQ}
\newblock
\APACjournalVolNumPages{Social Networks}{5}{2}{109--137}.
\PrintBackRefs{\CurrentBib}

\bibitem [\protect \citeauthoryear {%
Leicht%
\ \BBA {} Newman%
}{%
Leicht%
\ \BBA {} Newman%
}{%
{\protect \APACyear {2008}}%
}]{%
leicht2008community}
\APACinsertmetastar {%
leicht2008community}%
\begin{APACrefauthors}%
Leicht, E\BPBI A.%
\BCBT {}\ \BBA {} Newman, M\BPBI E.%
\end{APACrefauthors}%
\unskip\
\newblock
\APACrefYearMonthDay{2008}{}{}.
\newblock
{\BBOQ}\APACrefatitle {Community structure in directed networks} {Community
  structure in directed networks}.{\BBCQ}
\newblock
\APACjournalVolNumPages{Physical Review Letters}{100}{11}{118703}.
\PrintBackRefs{\CurrentBib}

\bibitem [\protect \citeauthoryear {%
Lyu%
\ \protect \BOthers {.}}{%
Lyu%
\ \protect \BOthers {.}}{%
{\protect \APACyear {2019}}%
}]{%
lyu2019integrative}
\APACinsertmetastar {%
lyu2019integrative}%
\begin{APACrefauthors}%
Lyu, Y.%
, Zauhar, R.%
, Dana, N.%
, Strang, C\BPBI E.%
, Wang, K.%
, Liu, S.%
\BDBL {}others%
\end{APACrefauthors}%
\unskip\
\newblock
\APACrefYearMonthDay{2019}{}{}.
\newblock
{\BBOQ}\APACrefatitle {Integrative single-cell and bulk RNA-seq analysis in
  human retina identified cell type-specific composition and gene expression
  changes for age-related macular degeneration} {Integrative single-cell and
  bulk rna-seq analysis in human retina identified cell type-specific
  composition and gene expression changes for age-related macular
  degeneration}.{\BBCQ}
\newblock
\APACjournalVolNumPages{bioRxiv}{}{}{768143}.
\PrintBackRefs{\CurrentBib}

\bibitem [\protect \citeauthoryear {%
Massen%
\ \BBA {} Doye%
}{%
Massen%
\ \BBA {} Doye%
}{%
{\protect \APACyear {2005}}%
}]{%
massen2005identifying}
\APACinsertmetastar {%
massen2005identifying}%
\begin{APACrefauthors}%
Massen, C\BPBI P.%
\BCBT {}\ \BBA {} Doye, J\BPBI P.%
\end{APACrefauthors}%
\unskip\
\newblock
\APACrefYearMonthDay{2005}{}{}.
\newblock
{\BBOQ}\APACrefatitle {Identifying communities within energy landscapes}
  {Identifying communities within energy landscapes}.{\BBCQ}
\newblock
\APACjournalVolNumPages{Physical Review E}{71}{4}{046101}.
\PrintBackRefs{\CurrentBib}

\bibitem [\protect \citeauthoryear {%
Matias%
\ \BBA {} Miele%
}{%
Matias%
\ \BBA {} Miele%
}{%
{\protect \APACyear {2015}}%
}]{%
matias2015statistical}
\APACinsertmetastar {%
matias2015statistical}%
\begin{APACrefauthors}%
Matias, C.%
\BCBT {}\ \BBA {} Miele, V.%
\end{APACrefauthors}%
\unskip\
\newblock
\APACrefYearMonthDay{2015}{}{}.
\newblock
{\BBOQ}\APACrefatitle {Statistical clustering of temporal networks through a
  dynamic stochastic block model} {Statistical clustering of temporal networks
  through a dynamic stochastic block model}.{\BBCQ}
\newblock
\APACjournalVolNumPages{arXiv preprint arXiv:1506.07464}{}{}{}.
\PrintBackRefs{\CurrentBib}

\bibitem [\protect \citeauthoryear {%
Medus%
, Acu{\~n}a%
\BCBL {}\ \BBA {} Dorso%
}{%
Medus%
\ \protect \BOthers {.}}{%
{\protect \APACyear {2005}}%
}]{%
medus2005detection}
\APACinsertmetastar {%
medus2005detection}%
\begin{APACrefauthors}%
Medus, A.%
, Acu{\~n}a, G.%
\BCBL {}\ \BBA {} Dorso, C\BPBI O.%
\end{APACrefauthors}%
\unskip\
\newblock
\APACrefYearMonthDay{2005}{}{}.
\newblock
{\BBOQ}\APACrefatitle {Detection of community structures in networks via global
  optimization} {Detection of community structures in networks via global
  optimization}.{\BBCQ}
\newblock
\APACjournalVolNumPages{Physica A: Statistical Mechanics and its
  Applications}{358}{2-4}{593--604}.
\PrintBackRefs{\CurrentBib}

\bibitem [\protect \citeauthoryear {%
Newman%
}{%
Newman%
}{%
{\protect \APACyear {2004}}%
}]{%
newman2004fast}
\APACinsertmetastar {%
newman2004fast}%
\begin{APACrefauthors}%
Newman, M\BPBI E.%
\end{APACrefauthors}%
\unskip\
\newblock
\APACrefYearMonthDay{2004}{}{}.
\newblock
{\BBOQ}\APACrefatitle {Fast algorithm for detecting community structure in
  networks} {Fast algorithm for detecting community structure in
  networks}.{\BBCQ}
\newblock
\APACjournalVolNumPages{Physical Review E}{69}{6}{066133}.
\PrintBackRefs{\CurrentBib}

\bibitem [\protect \citeauthoryear {%
Newman%
}{%
Newman%
}{%
{\protect \APACyear {2006}}%
}]{%
newman2006modularity}
\APACinsertmetastar {%
newman2006modularity}%
\begin{APACrefauthors}%
Newman, M\BPBI E.%
\end{APACrefauthors}%
\unskip\
\newblock
\APACrefYearMonthDay{2006}{}{}.
\newblock
{\BBOQ}\APACrefatitle {Modularity and community structure in networks}
  {Modularity and community structure in networks}.{\BBCQ}
\newblock
\APACjournalVolNumPages{Proceedings of the National Academy of
  Sciences}{103}{23}{8577--8582}.
\PrintBackRefs{\CurrentBib}

\bibitem [\protect \citeauthoryear {%
Newman%
\ \BBA {} Girvan%
}{%
Newman%
\ \BBA {} Girvan%
}{%
{\protect \APACyear {2004}}%
}]{%
newman2004finding}
\APACinsertmetastar {%
newman2004finding}%
\begin{APACrefauthors}%
Newman, M\BPBI E.%
\BCBT {}\ \BBA {} Girvan, M.%
\end{APACrefauthors}%
\unskip\
\newblock
\APACrefYearMonthDay{2004}{}{}.
\newblock
{\BBOQ}\APACrefatitle {Finding and evaluating community structure in networks}
  {Finding and evaluating community structure in networks}.{\BBCQ}
\newblock
\APACjournalVolNumPages{Physical Review E}{69}{2}{026113}.
\PrintBackRefs{\CurrentBib}

\bibitem [\protect \citeauthoryear {%
Pons%
\ \BBA {} Latapy%
}{%
Pons%
\ \BBA {} Latapy%
}{%
{\protect \APACyear {2005}}%
}]{%
pons2005computing}
\APACinsertmetastar {%
pons2005computing}%
\begin{APACrefauthors}%
Pons, P.%
\BCBT {}\ \BBA {} Latapy, M.%
\end{APACrefauthors}%
\unskip\
\newblock
\APACrefYearMonthDay{2005}{}{}.
\newblock
{\BBOQ}\APACrefatitle {Computing communities in large networks using random
  walks} {Computing communities in large networks using random walks}.{\BBCQ}
\newblock
\BIn{} \APACrefbtitle {International {S}ymposium on {C}omputer and
  {I}nformation {S}ciences} {International {S}ymposium on {C}omputer and
  {I}nformation {S}ciences}\ (\BPGS\ 284--293).
\PrintBackRefs{\CurrentBib}

\bibitem [\protect \citeauthoryear {%
Que%
, Checconi%
, Petrini%
\BCBL {}\ \BBA {} Gunnels%
}{%
Que%
\ \protect \BOthers {.}}{%
{\protect \APACyear {2015}}%
}]{%
que2015scalable}
\APACinsertmetastar {%
que2015scalable}%
\begin{APACrefauthors}%
Que, X.%
, Checconi, F.%
, Petrini, F.%
\BCBL {}\ \BBA {} Gunnels, J\BPBI A.%
\end{APACrefauthors}%
\unskip\
\newblock
\APACrefYearMonthDay{2015}{}{}.
\newblock
{\BBOQ}\APACrefatitle {Scalable community detection with the louvain algorithm}
  {Scalable community detection with the louvain algorithm}.{\BBCQ}
\newblock
\BIn{} \APACrefbtitle {2015 {IEEE} {I}nternational {P}arallel and {D}istributed
  {P}rocessing {S}ymposium} {2015 {IEEE} {I}nternational {P}arallel and
  {D}istributed {P}rocessing {S}ymposium}\ (\BPGS\ 28--37).
\PrintBackRefs{\CurrentBib}

\bibitem [\protect \citeauthoryear {%
Reichardt%
\ \BBA {} Bornholdt%
}{%
Reichardt%
\ \BBA {} Bornholdt%
}{%
{\protect \APACyear {2006}}%
}]{%
reichardt2006statistical}
\APACinsertmetastar {%
reichardt2006statistical}%
\begin{APACrefauthors}%
Reichardt, J.%
\BCBT {}\ \BBA {} Bornholdt, S.%
\end{APACrefauthors}%
\unskip\
\newblock
\APACrefYearMonthDay{2006}{}{}.
\newblock
{\BBOQ}\APACrefatitle {Statistical mechanics of community detection}
  {Statistical mechanics of community detection}.{\BBCQ}
\newblock
\APACjournalVolNumPages{Physical Review E}{74}{1}{016110}.
\PrintBackRefs{\CurrentBib}

\bibitem [\protect \citeauthoryear {%
Riedy%
, Meyerhenke%
, Ediger%
\BCBL {}\ \BBA {} Bader%
}{%
Riedy%
\ \protect \BOthers {.}}{%
{\protect \APACyear {2011}}%
}]{%
riedy2011parallel}
\APACinsertmetastar {%
riedy2011parallel}%
\begin{APACrefauthors}%
Riedy, E\BPBI J.%
, Meyerhenke, H.%
, Ediger, D.%
\BCBL {}\ \BBA {} Bader, D\BPBI A.%
\end{APACrefauthors}%
\unskip\
\newblock
\APACrefYearMonthDay{2011}{}{}.
\newblock
{\BBOQ}\APACrefatitle {Parallel community detection for massive graphs}
  {Parallel community detection for massive graphs}.{\BBCQ}
\newblock
\BIn{} \APACrefbtitle {{I}nternational {C}onference on {P}arallel {P}rocessing
  and {A}pplied {M}athematics} {{I}nternational {C}onference on {P}arallel
  {P}rocessing and {A}pplied {M}athematics}\ (\BPGS\ 286--296).
\PrintBackRefs{\CurrentBib}

\bibitem [\protect \citeauthoryear {%
Satija%
, Farrell%
, Gennert%
, Schier%
\BCBL {}\ \BBA {} Regev%
}{%
Satija%
\ \protect \BOthers {.}}{%
{\protect \APACyear {2015}}%
}]{%
satija2015spatial}
\APACinsertmetastar {%
satija2015spatial}%
\begin{APACrefauthors}%
Satija, R.%
, Farrell, J\BPBI A.%
, Gennert, D.%
, Schier, A\BPBI F.%
\BCBL {}\ \BBA {} Regev, A.%
\end{APACrefauthors}%
\unskip\
\newblock
\APACrefYearMonthDay{2015}{}{}.
\newblock
{\BBOQ}\APACrefatitle {Spatial reconstruction of single-cell gene expression
  data} {Spatial reconstruction of single-cell gene expression data}.{\BBCQ}
\newblock
\APACjournalVolNumPages{Nature Biotechnology}{33}{5}{495--502}.
\PrintBackRefs{\CurrentBib}

\bibitem [\protect \citeauthoryear {%
Shen%
\ \BBA {} Cheng%
}{%
Shen%
\ \BBA {} Cheng%
}{%
{\protect \APACyear {2010}}%
}]{%
shen2010spectral}
\APACinsertmetastar {%
shen2010spectral}%
\begin{APACrefauthors}%
Shen, H\BHBI W.%
\BCBT {}\ \BBA {} Cheng, X\BHBI Q.%
\end{APACrefauthors}%
\unskip\
\newblock
\APACrefYearMonthDay{2010}{}{}.
\newblock
{\BBOQ}\APACrefatitle {Spectral methods for the detection of network community
  structure: a comparative analysis} {Spectral methods for the detection of
  network community structure: a comparative analysis}.{\BBCQ}
\newblock
\APACjournalVolNumPages{Journal of Statistical Mechanics: Theory and
  Experiment}{2010}{10}{P10020}.
\PrintBackRefs{\CurrentBib}

\bibitem [\protect \citeauthoryear {%
Traag%
, Van~Dooren%
\BCBL {}\ \BBA {} Nesterov%
}{%
Traag%
\ \protect \BOthers {.}}{%
{\protect \APACyear {2011}}%
}]{%
traag2011narrow}
\APACinsertmetastar {%
traag2011narrow}%
\begin{APACrefauthors}%
Traag, V\BPBI A.%
, Van~Dooren, P.%
\BCBL {}\ \BBA {} Nesterov, Y.%
\end{APACrefauthors}%
\unskip\
\newblock
\APACrefYearMonthDay{2011}{}{}.
\newblock
{\BBOQ}\APACrefatitle {Narrow scope for resolution-limit-free community
  detection} {Narrow scope for resolution-limit-free community
  detection}.{\BBCQ}
\newblock
\APACjournalVolNumPages{Physical Review E}{84}{1}{016114}.
\PrintBackRefs{\CurrentBib}

\bibitem [\protect \citeauthoryear {%
Waltman%
, Van~Eck%
\BCBL {}\ \BBA {} Noyons%
}{%
Waltman%
\ \protect \BOthers {.}}{%
{\protect \APACyear {2010}}%
}]{%
waltman2010unified}
\APACinsertmetastar {%
waltman2010unified}%
\begin{APACrefauthors}%
Waltman, L.%
, Van~Eck, N\BPBI J.%
\BCBL {}\ \BBA {} Noyons, E\BPBI C.%
\end{APACrefauthors}%
\unskip\
\newblock
\APACrefYearMonthDay{2010}{}{}.
\newblock
{\BBOQ}\APACrefatitle {A unified approach to mapping and clustering of
  bibliometric networks} {A unified approach to mapping and clustering of
  bibliometric networks}.{\BBCQ}
\newblock
\APACjournalVolNumPages{Journal of Informetrics}{4}{4}{629--635}.
\PrintBackRefs{\CurrentBib}

\bibitem [\protect \citeauthoryear {%
White%
\ \BBA {} Smyth%
}{%
White%
\ \BBA {} Smyth%
}{%
{\protect \APACyear {2005}}%
}]{%
white2005spectral}
\APACinsertmetastar {%
white2005spectral}%
\begin{APACrefauthors}%
White, S.%
\BCBT {}\ \BBA {} Smyth, P.%
\end{APACrefauthors}%
\unskip\
\newblock
\APACrefYearMonthDay{2005}{}{}.
\newblock
{\BBOQ}\APACrefatitle {A spectral clustering approach to finding communities in
  graphs} {A spectral clustering approach to finding communities in
  graphs}.{\BBCQ}
\newblock
\BIn{} \APACrefbtitle {Proceedings of the 2005 {SIAM} {I}nternational
  {C}onference on {D}ata {M}ining} {Proceedings of the 2005 {SIAM}
  {I}nternational {C}onference on {D}ata {M}ining}\ (\BPGS\ 274--285).
\PrintBackRefs{\CurrentBib}

\bibitem [\protect \citeauthoryear {%
Yang%
, Algesheimer%
\BCBL {}\ \BBA {} Tessone%
}{%
Yang%
\ \protect \BOthers {.}}{%
{\protect \APACyear {2016}}%
}]{%
yang2016comparative}
\APACinsertmetastar {%
yang2016comparative}%
\begin{APACrefauthors}%
Yang, Z.%
, Algesheimer, R.%
\BCBL {}\ \BBA {} Tessone, C\BPBI J.%
\end{APACrefauthors}%
\unskip\
\newblock
\APACrefYearMonthDay{2016}{}{}.
\newblock
{\BBOQ}\APACrefatitle {A comparative analysis of community detection algorithms
  on artificial networks} {A comparative analysis of community detection
  algorithms on artificial networks}.{\BBCQ}
\newblock
\APACjournalVolNumPages{Scientific reports}{6}{1}{1--18}.
\PrintBackRefs{\CurrentBib}

\bibitem [\protect \citeauthoryear {%
G\BHBI Q.~Zhang%
, Wang%
\BCBL {}\ \BBA {} Li%
}{%
G\BHBI Q.~Zhang%
\ \protect \BOthers {.}}{%
{\protect \APACyear {2007}}%
}]{%
zhang2007enhancing}
\APACinsertmetastar {%
zhang2007enhancing}%
\begin{APACrefauthors}%
Zhang, G\BHBI Q.%
, Wang, D.%
\BCBL {}\ \BBA {} Li, G\BHBI J.%
\end{APACrefauthors}%
\unskip\
\newblock
\APACrefYearMonthDay{2007}{}{}.
\newblock
{\BBOQ}\APACrefatitle {Enhancing the transmission efficiency by edge deletion
  in scale-free networks} {Enhancing the transmission efficiency by edge
  deletion in scale-free networks}.{\BBCQ}
\newblock
\APACjournalVolNumPages{Physical Review E}{76}{1}{017101}.
\PrintBackRefs{\CurrentBib}

\bibitem [\protect \citeauthoryear {%
G\BHBI Q.~Zhang%
, Zhang%
, Yang%
, Cheng%
\BCBL {}\ \BBA {} Zhou%
}{%
G\BHBI Q.~Zhang%
\ \protect \BOthers {.}}{%
{\protect \APACyear {2008}}%
}]{%
zhang2008evolution}
\APACinsertmetastar {%
zhang2008evolution}%
\begin{APACrefauthors}%
Zhang, G\BHBI Q.%
, Zhang, G\BHBI Q.%
, Yang, Q\BHBI F.%
, Cheng, S\BHBI Q.%
\BCBL {}\ \BBA {} Zhou, T.%
\end{APACrefauthors}%
\unskip\
\newblock
\APACrefYearMonthDay{2008}{}{}.
\newblock
{\BBOQ}\APACrefatitle {Evolution of the Internet and its cores} {Evolution of
  the internet and its cores}.{\BBCQ}
\newblock
\APACjournalVolNumPages{New Journal of Physics}{10}{12}{123027}.
\PrintBackRefs{\CurrentBib}

\bibitem [\protect \citeauthoryear {%
J.~Zhang%
\ \BBA {} Chen%
}{%
J.~Zhang%
\ \BBA {} Chen%
}{%
{\protect \APACyear {2017}}%
}]{%
zhang2017hypothesis}
\APACinsertmetastar {%
zhang2017hypothesis}%
\begin{APACrefauthors}%
Zhang, J.%
\BCBT {}\ \BBA {} Chen, Y.%
\end{APACrefauthors}%
\unskip\
\newblock
\APACrefYearMonthDay{2017}{}{}.
\newblock
{\BBOQ}\APACrefatitle {A hypothesis testing framework for modularity based
  network community detection} {A hypothesis testing framework for modularity
  based network community detection}.{\BBCQ}
\newblock
\APACjournalVolNumPages{Statistica Sinica}{}{}{437--456}.
\PrintBackRefs{\CurrentBib}

\bibitem [\protect \citeauthoryear {%
S.~Zhang%
, Wang%
\BCBL {}\ \BBA {} Zhang%
}{%
S.~Zhang%
\ \protect \BOthers {.}}{%
{\protect \APACyear {2007}}%
}]{%
zhang2007identification}
\APACinsertmetastar {%
zhang2007identification}%
\begin{APACrefauthors}%
Zhang, S.%
, Wang, R\BHBI S.%
\BCBL {}\ \BBA {} Zhang, X\BHBI S.%
\end{APACrefauthors}%
\unskip\
\newblock
\APACrefYearMonthDay{2007}{}{}.
\newblock
{\BBOQ}\APACrefatitle {Identification of overlapping community structure in
  complex networks using fuzzy c-means clustering} {Identification of
  overlapping community structure in complex networks using fuzzy c-means
  clustering}.{\BBCQ}
\newblock
\APACjournalVolNumPages{Physica A: Statistical Mechanics and its
  Applications}{374}{1}{483--490}.
\PrintBackRefs{\CurrentBib}

\end{thebibliography}
\end{document}